\documentclass{PoS}

\usepackage{graphicx,amsmath,amssymb,fontenc,times,mathptmx}

\usepackage[utf8]{inputenc}

\usepackage{nicefrac}

\usepackage{tikz}
\usetikzlibrary{shadows}
\usetikzlibrary{arrows,shapes,positioning}
\usetikzlibrary{decorations.markings}
\usetikzlibrary{decorations.pathmorphing}
\usetikzlibrary{decorations.pathreplacing}
\tikzstyle arrowstyle=[scale=1]
\tikzstyle directed=[postaction={decorate,decoration={markings, mark=at position .65 with {\arrow[arrowstyle]{stealth}}}}]
\tikzstyle end directed=[postaction={decorate,decoration={markings, mark=at position 1 with {\arrow[arrowstyle]{stealth}}}}]
\tikzstyle reverse directed=[postaction={decorate,decoration={markings, mark=at position .65 with {\arrowreversed[arrowstyle]{stealth};}}}]
\tikzstyle{ann} = [fill=white,font=\footnotesize,inner sep=1pt]

\title{Parton Distributions from Lattice QCD with Momentum
Smearing}

\ShortTitle{PDFs from Lattice QCD}

\author{Constantia Alexandrou$^{ab}$,
  Krzysztof Cichy$^{cd}$,
        Martha Constantinou$^{e}$,  
	Kyriakos Hadjiyiannakou$^{a}$, 
	Karl Jansen$^{f}$, 
	Fernanda Steffens$^{f}$,
	\speaker{Christian Wiese}$\;^{f}$
\\
\\
        \llap{$^a$}Department of Physics, University of Cyprus, P.O. Box 20537, 1678 Nicosia, Cyprus\\
	\llap{$^b$}Computation-based Science and Technology
	Research Center, Cyprus Institute, 20 Kavafi Str.,
	Nicosia 2121, Cyprus\\
	\llap{$^c$}Goethe-Universität Frankfurt am Main,
	Institut für Theoretische Physik,
	Max-von-Laue-Strasse~1, 60438 Frankfurt am Main,
	Germany\\
	\llap{$^d$}Faculty of Physics, Adam Mickiewicz
	University, Umultowska 85, 61-614 Poznań, Poland\\
	\llap{$^e$}Temple University, 1925 N. 12th Street,
	Philadelphia, PA 19122, USA\\
	\llap{$^f$}John von Neumann Institute for Computing (NIC), DESY, Platanenallee 6, D-15738 Zeuthen, Germany\\
	\\
        E-mail: \email{christian.wiese@desy.de}}

\abstract{ 

In this work we continue our effort to explore a recent
proposal, which allows light-cone distributions to be
extracted from purely spatial correlations, being thus
accessible to lattice methods. In order to test the
feasibility of this method, we present our latest results 
from a twisted mass lattice calculation of the flavor non-singlet momentum,
helicity and transversity distributions of the nucleon. 
Furthermore, we apply a newly
proposed momentum improved smearing, which has the potential
to reach higher nucleon momenta as required for a safe
matching procedure to the physical distribution functions.}

\FullConference{34th annual International Symposium on Lattice Field Theory\\
		24-30 July 2016\\
		University of Southampton, UK}

\begin{document}

\section{Introduction}

Although parton distribution functions are the
fundamental objects describing the inner structure of
hadrons, they were so far not calculated from first
principles. In the past, lattice QCD has successfully been
employed for the computation of hadronic spectra and form
factors, for instance. Yet calculations of quark distributions are still
missing, since they are given by light-cone correlation
functions and light-like distances are not accessible on an
Euclidean lattice.

A new method to deal with this problem was proposed in
Ref.\,\cite{Ji:2013dva} and employs the computation of a
purely spatial quasi-distribution in
a finite momentum frame. How to relate this
quasi-distribution to the physical PDF and general studies
of this method have already been addressed in a handful of
papers, e.g. \cite{Xiong:2013bka, Lin:2014zya, Alexandrou:2015rja,
Chen:2016utp, Ishikawa:2016znu, Chen:2016fxx}.

A crucial point when making the connection to PDFs is a
large momentum limit for the nucleon boost. On the lattice,
this provides a challenge, since large momentum nucleon
observables are known to be very noisy. Thus, we study the
effect of including a momentum dependent quark field
smearing \cite{Bali:2016lva} into our calculation, in order
to improve the quality of the signal, especially for large
nucleon momenta.

A comprehensive work on the calculation of PDFs with more
elaborate studies of the recent developments and our newest
results can be found in our latest paper
\cite{Alexandrou:2016jqi}.

\section{Lattice calculation \& momentum smearing}

The quasi-distributions are computed from
\begin{equation}
\tilde{q}(x, \Lambda, P_3) = \int_{-\infty}^\infty
\frac{dz}{4\pi} e^{-izk_3} \langle P
|\bar{\psi}(z)\gamma_3 W_3(z,0) \psi(0) |P\rangle\,,
\label{EQN_PDF_FT}
\end{equation}
where $W_j(z,0)$ is the Wilson line from $0$ to $z$ in the
spatial $j$ direction, $k_3=xP_3$ and the Euclidean momentum
is $P=(0, 0, P_3, P_4)$. It is required that the Wilson line and the spatial nucleon
momentum boost point into the same direction. 
In lattice QCD, one can compute matrix elements of operator
as suitable ratios of three- and two-point correlation
functions. 
The full set of PDFs can be accessed by using three-point
functions where the inserted operators have
the Dirac structures 
\begin{itemize}  
  \item $\gamma_3$, for the case of the unpolarized momentum
    quasi-distributions $\tilde{q}(x,\Lambda, P_3)$\,;
  \item $\gamma_3 \gamma_5$, for the case of the helicity
    quasi-distributions $\Delta \tilde{q}(x,\Lambda,P_3)$\,; 
  \item $\gamma_3 \gamma_j$ ($j=1,2$), for the case of the transversity
    quasi-distributions $\delta \tilde{q}(x,\Lambda,P_3)$\,.
  \end{itemize} 
Due to the rotational invariance on the lattice, one is
certainly not restricted to the 3-direction and can easily
generalize the operators to the other two
directions.
In order to obtain a quasi-distribution from the computed
matrix elements a Fourier transformation has to be
performed. The quasi-distribution can be related to the physical
PDF by a one-loop matching and a mass correction. Details
for these steps can be found in
Refs.\,\cite{Alexandrou:2015rja,Chen:2016utp}, for example.

In former studies we applied standard Gaussian smearing to
the nucleon fields in order to improve the overlap to the
nucleon ground state. This, however, is only valid for
momentum zero and deteriorates the signal for large momenta. 
Thus, in Ref.\,\cite{Bali:2016lva} a new type of smearing, the
momentum smearing, is proposed. This smearing is constructed
in a way to increase the overlap of the used nucleon field
with a momentum boosted nucleon ground state.
In contrast to the Gaussian
smearing quark fields are smeared according to
\begin{equation}
  S_{mom}\psi(x) = \frac{1}{1+6\alpha}(\psi(x) + \alpha
  \sum_j U_j(x)e^{ik\hat{j}}\psi(x+\hat{j}))\,,
\end{equation}
where $k=\zeta P$, with $P$ the lattice momentum of the
nucleon and $\zeta$ a tunable parameter. This form looks
very similar to the definition of Gaussian smearing with an
addition of an exponential factor multiplying the gauge
links in the direction of the momentum boost. In practice, we
use a Gaussian smearing routine with the standard nucleon
parameters (50 steps, $\alpha=4$) and include a gauge field
with a complex phase $\tilde{U}_j(x) = U_j(x)e^{ik\hat{j}}$. For now we follow \cite{Bali:2016lva}
and choose $\zeta$ to be 0.45.

All calculations presented here were performed on an ETMC
(European Twisted Mass Collaboration) gauge field ensemble
\cite{Baron:2010bv}, with $N_f = 2+1+1$ flavors of
maximally twisted mass fermions and a volume of $32^3 \times
64$. The bare coupling is set to $\beta = 1.95$,
corresponding to a lattice spacing of $a \approx 0.082$\,fm.
The twisted mass parameter is $a\mu = 0.0055$, which gives a
pion mass of $m_{PS}\approx 370$\,MeV. 
Because we still do not have a proper renormalization of the
involved operator we apply 5 steps of HYP smearing
\cite{Hasenfratz:2001hp} to the gauge links in the Wilson
line. This is expected to bring the values of the
renormalization constants close to their tree-level values.

For the computation of
the matrix elements we used only 50 gauge
configurations with 3 sequential propagators (one for each
spatial direction), resulting in 150 measurements. In case
of the unpolarized momentum distribution and $P_3 = 10 \pi / L$, we
used 100 gauge configurations. Currently the source-sink
separation is set to $t_s=8a$, in order to get a better
signal. Larger separations and thus the influence of excited
states will be studied in the future.

\section{Results}

\begin{figure}
  \includegraphics{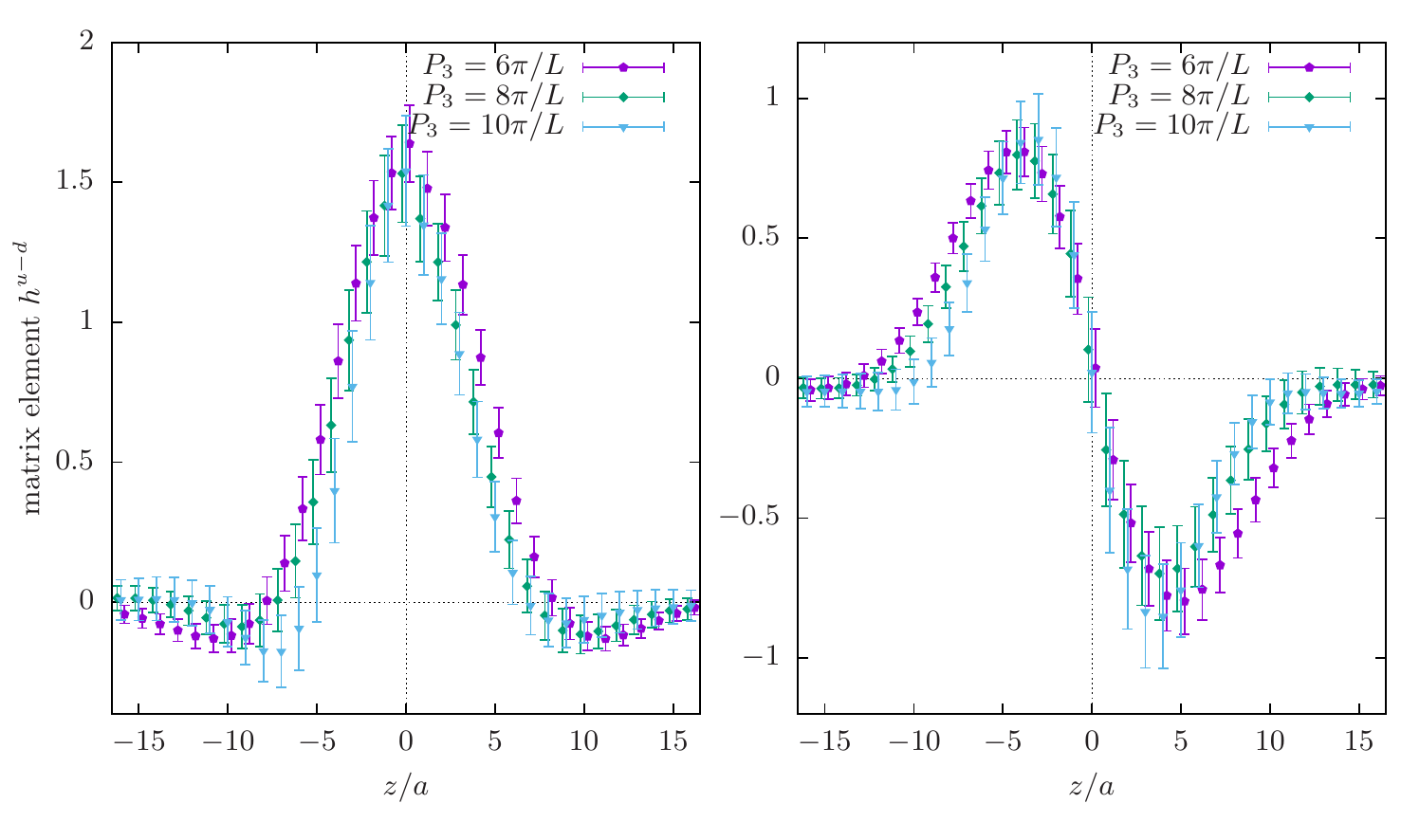}
  \caption{\label{FIG_ME_UP}Unrenormalized matrix elements of the vector
  operator (for the momentum distribution) and three
  different nucleon boost momenta, {\bf left:}
real part, {\bf right:} imaginary part.}
\end{figure}

\begin{figure}
  \includegraphics{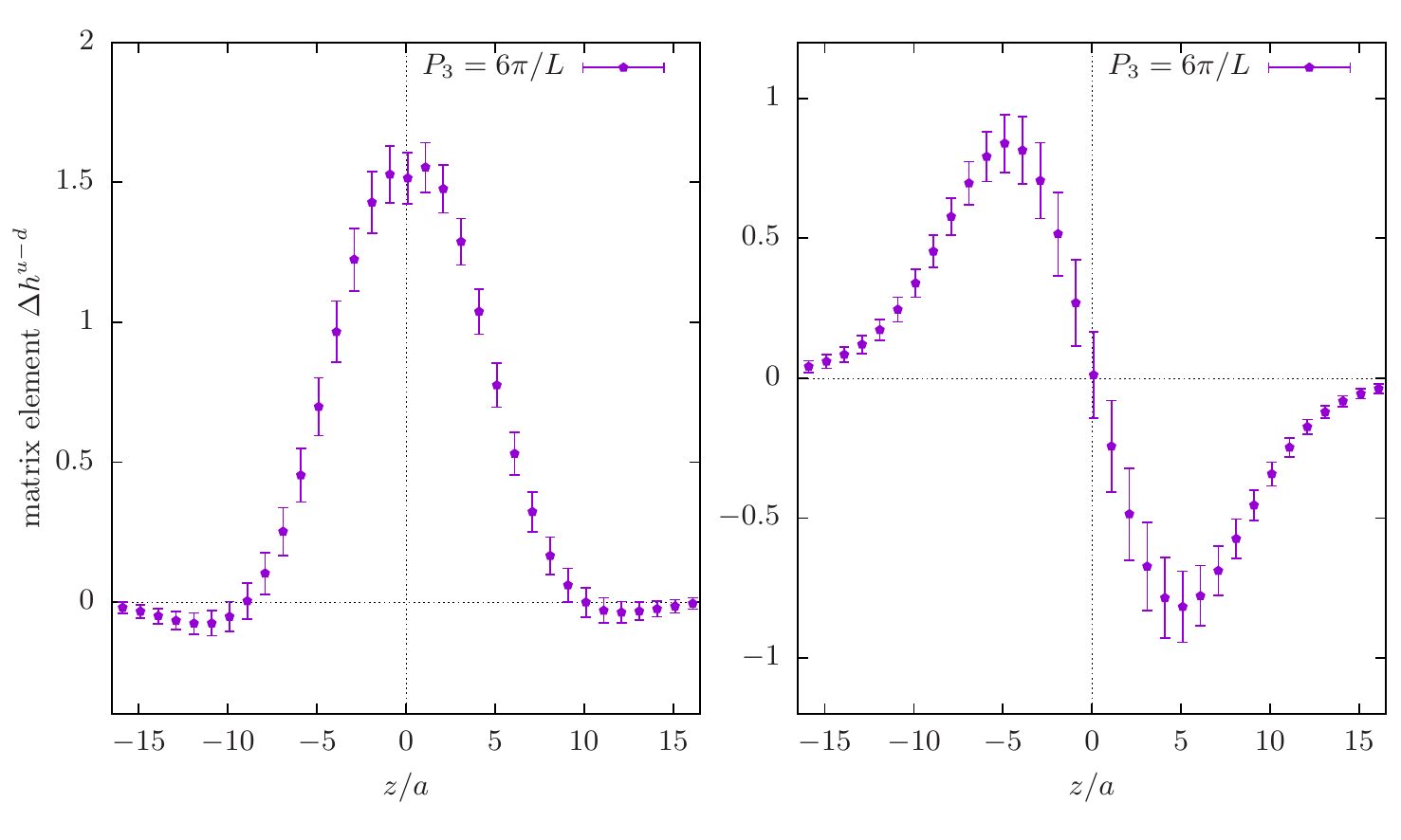}
  \caption{\label{FIG_ME_HEL}Unrenormalized matrix elements of the
  axial-vector operator (for the helicity distribution) and
$P_3 = 6\pi/L$, {\bf left:} real part, {\bf right:}
imaginary part.}
\end{figure}

\begin{figure}
  \includegraphics{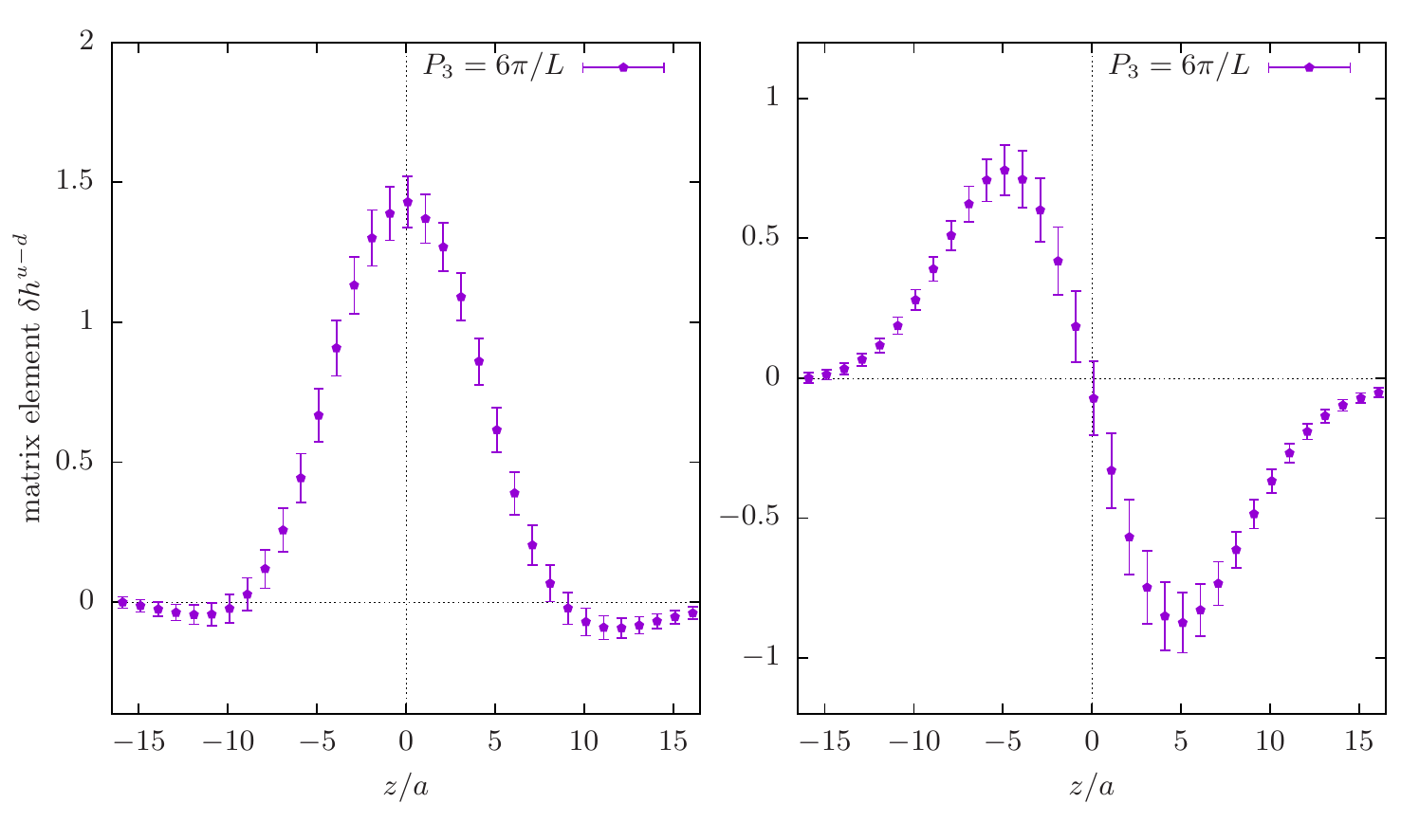}
  \caption{\label{FIG_ME_TRA}Unrenormalized matrix elements of the
  tensor operator (for the transversity distribution) and
$P_3 = 6\pi/L$, {\bf left:} real part, {\bf right:}
imaginary part.}
\end{figure}

In order to show the quality of our lattice calculation, we
show the unrenormalized matrix elements for the momentum,
the helicity and transversity cases in
Figs.\,\ref{FIG_ME_UP}, \ref{FIG_ME_HEL} and
\ref{FIG_ME_TRA}. To demonstrate the capabilities of the
momentum smearing we show results for momenta up to
$P_3=10\pi/L$ for the unpolarized case. When comparing to
previous results that employ the stochastic method
\cite{Alexandrou:2016jqi} one can see that a factor of
$\approx 200$ less measurements is necessary to obtain
results with a comparable uncertainty for a single momentum.

\begin{figure}
  \includegraphics{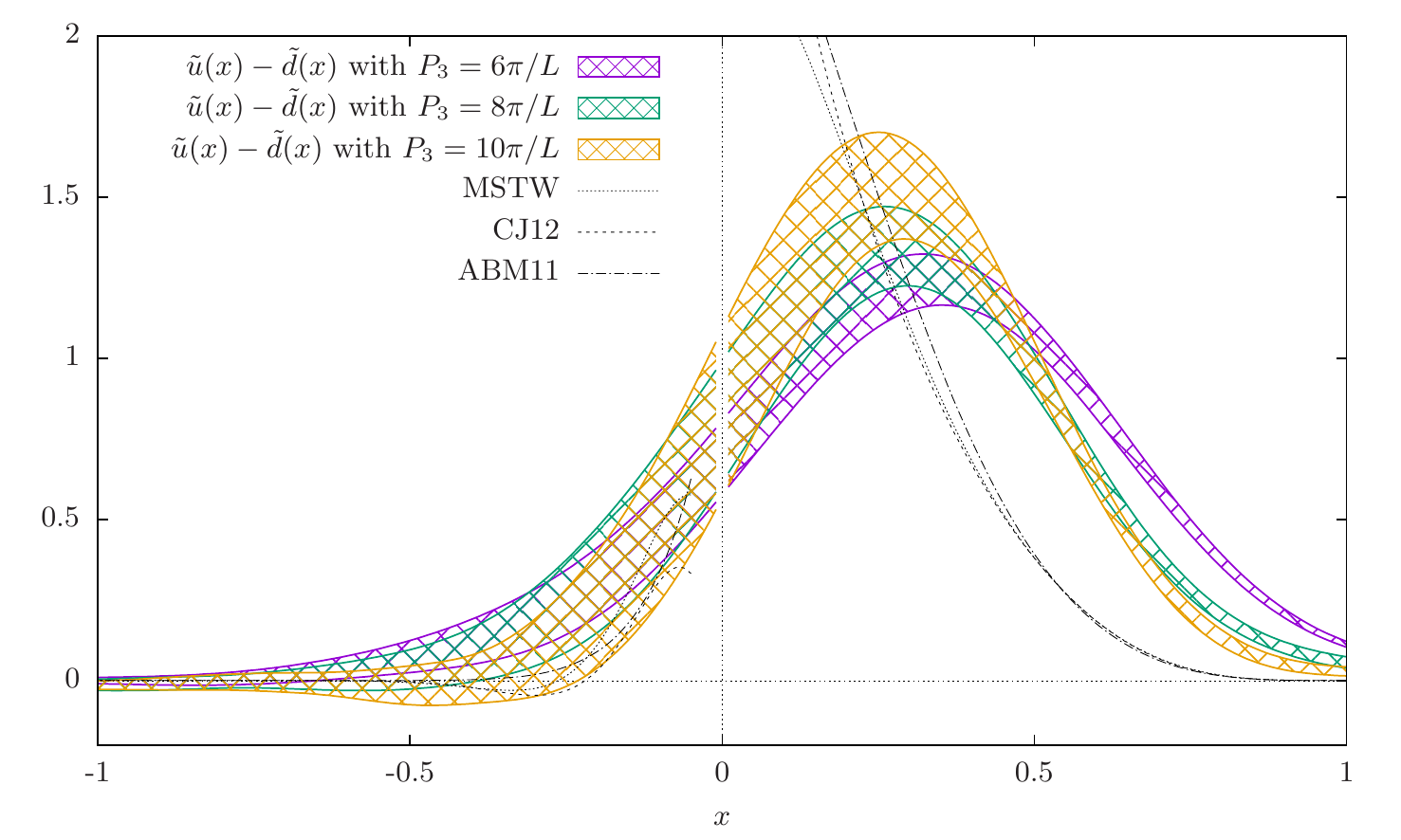}
  \caption{\label{FIG_PDF_MOM}Momentum quasi-distribution for
    three different nucleon momenta. MSTW
    \cite{Martin:2009iq}, CJ12 \cite{Owens:2012bv} and
    ABM11 \cite{Alekhin:2012ig} are
phenomenologically extracted distributions plotted for
orientation.}
\end{figure}

\begin{figure}
  \includegraphics{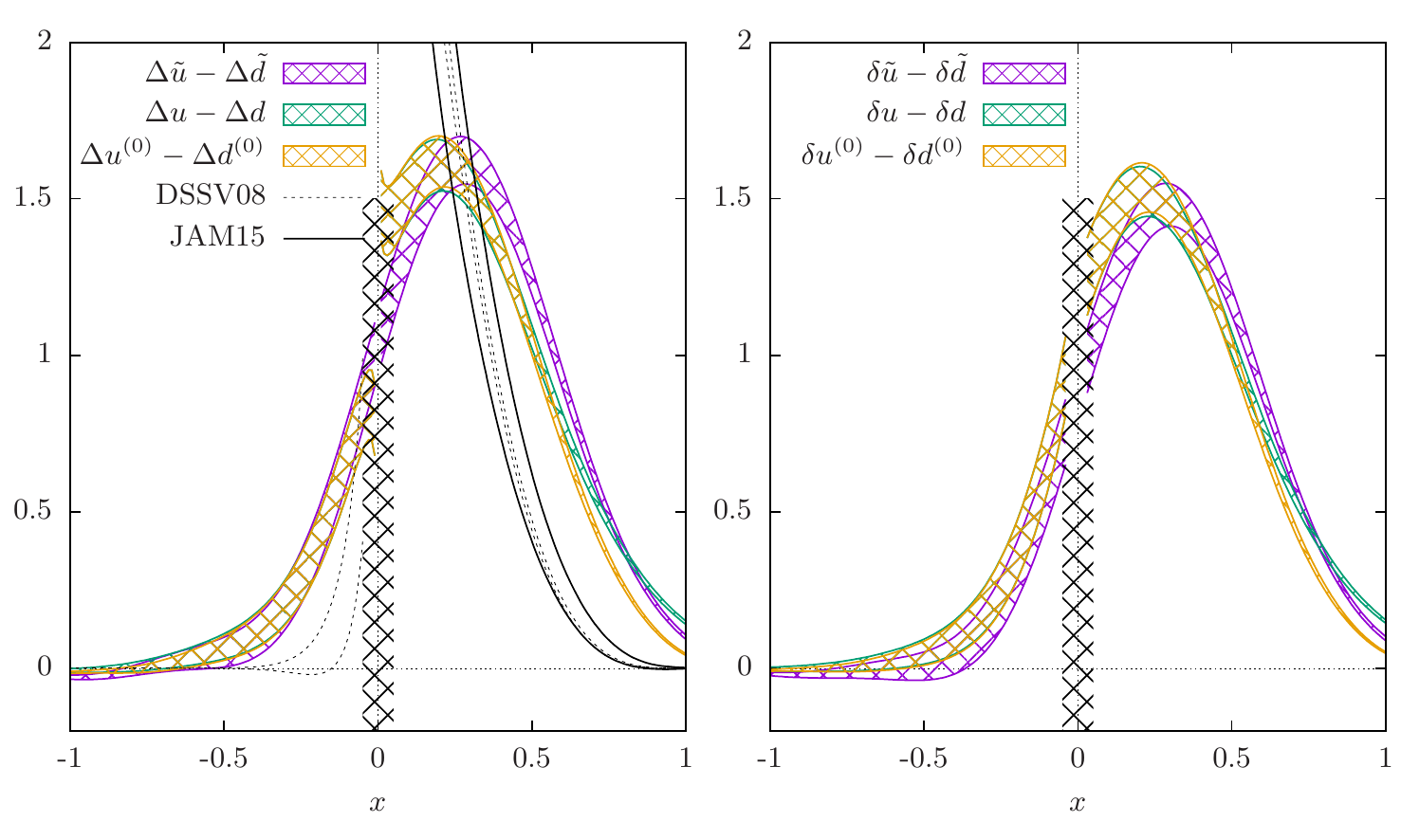}
  \caption{\label{FIG_PDF_HELTRA}Quasi-distribution, finite
    nucleon mass distribution and PDF after nucleon mass
    corrections for $P_3=6\pi/L$, {\bf left:} helicity, {\bf
    right:} transversity. DSSV08 \cite{deFlorian:2009vb} and
    JAM15 \cite{Sato:2016tuz} are
phenomenologically extracted distributions plotted for
orientation.}
\end{figure}

Consequently, in Fig.\,\ref{FIG_PDF_MOM} we show the momentum
dependence of the unpolarized momentum quasi-distribution. One can
see that for larger momenta there is a trend that
points into the direction of the phenomenological PDF
curves. However, in order to make proper statements here we
will need probably even larger momenta and a procedure to
extrapolate to very large or even infinite momentum. We show
only the quasi-distributions in this plot, since the
influence of the matching and mass correction is very small
for momenta larger than $P_3=6\pi/L$. This can especially
be seen when looking at the results for the helicity and
transversity distributions for $P_3=6\pi/L$ in
Fig.\,\ref{FIG_PDF_HELTRA}. Here, the difference between the
quasi-distribution, the finite mass distribution and the
physical PDF is almost negligible. For even larger momenta
there should be no visible difference.

\section{Conclusion and Outlook}
In this proceeding we gave a brief update on our current
effort to compute parton distribution functions of the
nucleon with lattice QCD methods. We especially focused on
the recent inclusion of momentum dependent smearing. We
were able to show that by using this smearing in combination
with the sequential method it is possible to reduce the
error for larger momenta up to a factor of 200 in comparison
to our previous work where we used the stochastic method. We
applied this new smearing to the momentum, the helicity
and the transversity distribution.

In the future, the application of momentum smearing will
thus enable us to reach sufficiently large momenta in order
to make a connection to the physical light cone
distribution. When other systematic effects, e.g. the quark
masses and cut-off effects and, in particular, the question
of renormalization, are under control, we will be
able to tell, if PDFs can be extracted from a lattice QCD
calculation.

\section*{Acknowledgements}
K.C. was supported in part
by the Deutsche Forschungsgemeinschaft (DFG), project nr.
CI 236/1-1. F.S. was partly supported by CNPq contract number 249168/2013-8.
We thank our fellow members of ETMC for their
constant collaboration. In particular
helpful discussions with G.C. Rossi are gratefully
acknowledged. We are grateful
to the John von Neumann Institute for Computing (NIC), the
Jülich Supercomputing
Center and the DESY Zeuthen Computing Center for their
computing resources and
support. This work has been supported by the Cyprus Research
Promotion Foundation through the  Project Cy-Tera (NEA
Y$\Pi$O$\Delta$OMH/$\Sigma$TPATH/0308/31) co-financed by the
European Regional Development Fund. This work has 
received funding from the European Union’s Horizon 2020
research and 
innovation programme under the Marie Sklodowska-Curie grant
agreement No 642069 (HPC-LEAP).

\bibliographystyle{JHEP}
\bibliography{references.bib}

\end{document}